\documentclass[]{aastex631}

\shorttitle{BBN Initial Conditions}
\shortauthors{Sharpe, Lewis \& Barnes}
\graphicspath{{./}{figures/}}

\begin{document}

\title{Big Bang Nucleosynthesis Initial Conditions: Revisiting \citet{1967ApJ...148....3W} }

\correspondingauthor{Geraint F. Lewis}
\email{geraint.lewis@sydney.edu.au}

\author{Charlie Sharpe}
\affiliation{Sydney Institute for Astronomy, School of Physics, A28, \\
The University of Sydney, NSW 2006, Australia}

\author[0000-0003-3081-9319]{Geraint F. Lewis}
\affiliation{Sydney Institute for Astronomy, School of Physics, A28, \\
The University of Sydney, NSW 2006, Australia}

\author{Luke A. Barnes}
\affiliation{Western Sydney University, Locked Bag 1797, \\ 
Penrith South, NSW 2751, Australia}



\begin{abstract}
We revisit \citet{1967ApJ...148....3W}, a classic contribution in the development of Big Bang Nucleosynthesis.
We demonstrate that it presents an 
incorrect expression for the temperature of the early universe as a function of time in the high temperature limit, $T \gtrsim 10^{10}$K. As this incorrect expression has been reproduced elsewhere, we present a corrected form for the initial conditions required for calculating the formation of the primordial elements in the Big Bang.
\end{abstract}

\keywords{Cosmology (343)}

\section{Introduction} \label{sec:intro}
Big Bang Nucleosynthesis (BBN) is considered an outstanding success and pillar of modern cosmology, explaining the abundance of primordial elements within the universe
\citep{Tytler_2000,barnes_lewis_2020}. Calculating the output of BBN requires integrating a series of coupled differential equations representing the sources and sinks of various elements, encompassing cross-sections that are temperature dependent. Such integrals, of course, depend upon the initial conditions. 

The structure of this research note is as follows; in Section \ref{sec: BBN IC}, we discuss the process \citet{1967ApJ...148....3W} take in deriving the initial conditions for the integration of BBN in the early universe. We then discuss the issue with their results in Section \ref{sec: Correct IC} before concluding in Section \ref{sec: Conclusions}.

\section{BBN Initial Conditions}\label{sec: BBN IC}
In their classic paper, \citet{1967ApJ...148....3W} derive the dependence of the temperature of the early universe on time, in the high temperature limit $T \gtrsim 10^{10}$K. To do this, they first derive the work-energy equation for the Friedmann–Lemaître–Robertson–Walker metric. The baryon mass density $\rho_b$ scales as $R^{-3}$ and the neutrino energy density $\rho_\nu$ scales as $R^{-4}$, where $R$ is the scale factor of the universe. After some algebra, they arrive at the following first-order differential equation for the expansion $R(T)$,
\begin{equation}
    \frac{d R}{d T}=\frac{-R}{3\left[\rho_{1}(T)+p_{1}(T) / c^{2}\right]} \frac{d\rho_1(T)}{dT}\label{eq: A13 of Wagoner}
\end{equation}
where $\rho_1 = \rho_e + \rho_\gamma$ is the sum of the energy densities of electrons and radiation, and $p_1 = p_e + p_\gamma$ is the sum of the pressures of electrons and radiation. They then combine this equation with the Friedmann equation to arrive at a differential equation for $T(t)$,
\begin{equation}
    \frac{d T}{d t} = \mp(24 \pi G \rho)^{1 / 2}\left[\rho_{1}(T)+\frac{p_{1}(T)}{c^{2}}\right]\left[\frac{d \rho_{1}(T)}{d T}\right]^{-1} \label{eq: A14 from Wagoner}
\end{equation}
Now, at such high temperatures ($T \gtrsim 10^{10}$K), electrons are highly relativistic with energy density $\frac{7}{4} \rho_\gamma$, the same as neutrinos. The coefficient of $\frac{7}{4}$ comes from the number of species and allowed spin states of each respective particle. This gives $\rho = \frac{9}{2}\rho_\gamma$ and $\rho_1 = \frac{11}{4}\rho_\gamma$. Hence, assuming an equation of state of $\frac{1}{3}$ for highly relativistic species, they finally arrive at a relationship between the temperature of the universe and time (their equation A15),
\begin{equation} 
    T_9 =  ( 12 \pi G a_{r}c^{-2} )^{1/4}\ t^{-1/2} = 10.4\ t^{-1/2} \label{eq: A15 from wagoner}
\end{equation}
Here, $T_9$ is the temperature, $G$ is Newton's gravitational constant, $a_r = 4\sigma/c$ is the radiation density constant, $\sigma = 2\pi^2 k_b^2/15h^3c^2$ is the Stefan-Boltzmann constant, and $c$ is the speed of light. The right-hand expression gives the temperature of the universe in units of $10^9$K when the time, $t$, is measured in seconds from $t=0$.

There are some interesting observations to be made about Equation \ref{eq: A15 from wagoner}, but perhaps the most important is that while the second equality, $T_9 = 10.4t^{-1/2}$, is correct, the first equality is \textit{incorrect}. 
Most obviously, the dimensions are incorrect. Working in cgs units, the central expression in Equation \ref{eq: A15 from wagoner} has units of s$^{-1}$K$^{-1}$, which is clearly inconsistent with $T_9$ having units of kelvin.

\section{Correcting the Initial Conditions}\label{sec: Correct IC}
\citet{1967ApJ...148....3W} state that  details that are relevant to their derivation of Equation \ref{eq: A15 from wagoner}  are provided by \citet{1953PhRv...92.1347A}. However, this earlier paper finds that temperature follows $T_9 = 15.2t^{-1/2}$, which has a different numerical factor from Equation \ref{eq: A15 from wagoner}. This discrepancy arises from the fact that \citet{1953PhRv...92.1347A} do not consider the presence of electrons or neutrinos at early times, and hence use $\rho = \rho_1 = \rho_\gamma$.

Directly solving Equation \ref{eq: A14 from Wagoner} we find that, for time in seconds, 
\begin{equation}
    T_9 = (48\pi a_{r}c^{-2}G)^{-1/4}t^{-1/2} = 10.4 t^{-1/2} \label{eq: Correct units}
\end{equation}
This expression is dimensionally correct, with the central equation having units K, and yields the same numerical factor as presented in  \citet{1967ApJ...148....3W}.

\section{Conclusions}\label{sec: Conclusions}
Normally, such a minor typographical error in a paper has no significant consequences and so can pass unnoted. 
However, the incorrect expression presented in \citet{1967ApJ...148....3W} has been propagated into other key publications, and numerical codes for calculating BBN. We highlight two of these below;
\begin{itemize}
\item \citet{1992STIN...9225163K} presents a fortran implementation of the integration of the coupled BBN equations named {\tt NUC123}\footnote{{\tt https://github.com/ckald/KAWANO-sterile}}. This directly reproduces the expression from \cite{1967ApJ...148....3W} as the initial condition for the integration (their equation D.6). However, an examination of the fortran source code reveals that the initial time is set to be {\tt t   = 1/(const1*t9)**2       !Initial time (Ref 1)} where {\tt PARAMETER (const1=0.09615)   !Relation between time and temperature}, so only the numerical aspect of Equation \ref{eq: A15 from wagoner} is employed.
    \item \citet{2020CoPhC.24806982A} presents a $C$ implementation of BBN known as {\tt AlterBBN}\footnote{{\tt https://alterbbn.hepforge.org}}. They too cite Equation \ref{eq: A15 from wagoner} in determining the initial conditions of their integration. Exploring the source code, the initial time is set to {\tt double t=sqrt(12.*pi*G*sigma\_SB)/pow(Ti,2.); } and so directly implements the incorrect form of the initial time. We note that in the natural (energy) units employed in {\tt AlterBBN} results in a very early starting time, well into the high temperature limit, and so does not impact the resulting integration. 
    We have discussed this with the {\tt AlterBBN} author(s) and the initial conditions will be corrected in a forthcoming update (Arbey 2021, priv. comm.).
\end{itemize}

\begin{acknowledgments}
This work was undertaken as part of Charlie Sharpe's honours project at the University of Sydney's School of Physics.
\end{acknowledgments}



\bibliography{paper}{}
\bibliographystyle{aasjournal}


\end{document}